\documentstyle[onecolumn]{mn}
\input psfig
\def\ltsim{\lower.5ex\hbox{$\; \buildrel < \over \sim \;$}}
\def\gtsim{\lower.5ex\hbox{$\; \buildrel > \over \sim \;$}}
\def\ltsim{\lower.5ex\hbox{$\; \buildrel < \over \sim \;$}}
\def\gtsim{\lower.5ex\hbox{$\; \buildrel > \over \sim \;$}}
\def\pp{\noindent\parshape 2 0truecm 15truecm 2truecm 13truecm}
\def\apj#1;#2;#3;#4 {\par\pp#1, {\it #2\/}, {\bf #3}, #4. \par}
\def\$${$$}

\def\bfr{{\bf r}}

\def\j3s{$J_3(s)$}
\def\j3r{$J_3(r)$}

\def\kms{\ifmmode {\rm \ km \ s^{-1}}\else $\rm km \ s^{-1}$\fi}

\def\ltsima{$\; \buildrel < \over \sim \;$}
\def\simlt{\lower.5ex\hbox{\ltsima}}
\def\gtsima{$\; \buildrel > \over \sim \;$}
\def\simgt{\lower.5ex\hbox{\gtsima}}

\def\al2{{$a_l^2$}}
\def\plcos{\ifmmode {P_l(\cos(\theta)}\else $P_l(\cos(\theta)$\fi}

\def\ltsim{\lower.5ex\hbox{$\; \buildrel < \over \sim \;$}}
\def\gtsim{\lower.5ex\hbox{$\; \buildrel > \over \sim \;$}}
\def\simprop{\lower.5ex\hbox{$\; \buildrel \propto \over \sim \;$}}

\def\U0{{ {\bf V}_{obs}\cdot\hat\bfr}}

\def\cN{{\cal N}}

\title[Dark matter halos]{ Mass growth and density profiles of dark
matter halos in hierarchical clustering}

\author[Nusser \& Sheth]{Adi Nusser \& Ravi K. Sheth \\
Max-Planck Institute f\"ur Astrophysik \\
Karl-Schwarzschild-Str. 1\\
85740 Garching, Germany}

\begin{document}

\maketitle

\begin{abstract}
We develop a model for the growth of dark matter halos and  
use it to study their evolved density profiles.
In this model, halos are spherical and form by
quiescent accretion of matter in clumps, called satellites. 
The halo mass as a function of redshift is given by the mass of 
the most massive progenitor, and is determined from Monte-Carlo 
realizations of the merger-history tree.   
Inside the halo, satellites move under the action of the 
gravitational force of the halo and a dynamical friction drag force.  
The associated equation of motion is solved numerically.  
The energy lost to dynamical friction is transferred to the halo 
in the form of kinetic energy.  
As they sink into the halo, satellites continually lose matter as 
a result of tidal stripping.  The stripped matter moves inside the 
halo free of dynamical friction.  
The evolved density profiles are steeper than those obtained by 
assuming that, once they have been accreted onto the parent halo, 
satellites remain at a fixed distance from the halo center.  
We find that the final density profile depends mainly on the rate 
of infall of matter onto the halo.  This, in turn, depends on the 
initial fluctuation field as well as on cosmology.
For mass scales where the effective spectral index of the initial 
density field is less than $-1$, the model predicts a profile which 
can only approximately be matched by the one parameter family of curves 
suggested by Navarro, Frenk and White (1997).  For scale-free 
power-spectra with initial slope $n$, the density 
profile within about 1\% of the virial radius is 
$\rho\propto r^{-\beta}$, with $3(3+n)/(5+n)\le\beta\le 3(3+n)/(4+n)$.

\end{abstract}

\begin{keywords}
 cosmology: dark matter-galaxy formation
\end{keywords}

\section{introduction}

A long standing question in the dynamics of cosmological gravitating
systems is: how well do dynamically evolved systems retain memory of
their initial conditions?  On large scales, where the evolution is
still in the the weakly non-linear regime, the growing mode of the
initial density fluctuations can, in principle, be fully recovered, if
the present velocity or density field is given ({\it e.g.,} Peebles
1989, Nusser \& Dekel 1992).  On small scales where shell crossing
has occurred and ``virialised'' objects (halos) have formed, the
situation is less clear.  Using techniques of statistical mechanics,
Lynden-Bell (1967) showed that starting from a general initial state,
a gravitating system could, via a process he termed ``violent
relaxation'', reach a quasi-equilibrium state which is almost
independent of the initial conditions. Therefore, under the restricted
conditions he assumed in his analysis, a gravitating system develops 
a ``universal'' density profile independent of the initial conditions.

The assumptions underlying Lynden-Bell's analysis are hard to
justify in the hierarchical scenario for the formation of structure in
an expanding universe.  Nevertheless,  high resolution cosmological 
N-body simulations of the gravitational clustering of collisionless 
particles from hierarchical initial conditions strongly suggest 
that dark matter halos do indeed develop a universal final density 
profile (e.g. Dubinski \& Calberg 1991, Lemson 1995, 
Navarro Frenk \& White 1995, 1996, Cole \& Lacey 1996, 
Moore et. al. 1997).  Navarro, Frenk \& White (1995, 1996; hereafter NFW) 
found that density profiles of halos of different masses in a variety 
of cosmological models could be fit with the following one parameter 
functional fit
\begin{equation}
\frac{\rho_{\rm N}(r)} {\rho_b}=\frac{\delta_N}{ \frac{r}{r_{\rm v}}\left[
1+c \frac{r}{r_{\rm v}}\right]^2} ,
\label{nfw}
\end{equation}
where $\rho_b$ is the background density and $r_{\rm v}$ is the virial
radius of the halo, defined as the radius within which the average
density is 178 times that of the background. With this definition
for $r_{\rm v}$, the parameter $\delta_N$ can be expressed in terms of 
the concentration parameter $c$; thus, $c$ is the only free
parameter in the fit (\ref{nfw}). It has been argued by NFW that 
$c$ is directly related to the formation time of a given halo. 
The claim that density profiles could be fitted with a one parameter 
functional form, has recently been challenged by Klypin et. al. (1998) 
who found, using N-body simulations of CDM-like models,
that the scatter about a one parameter fit is substantial, 
indicating that the structure of halo density profiles 
involves more than just one physical parameter.  

Because of the lack of a general analytic technique for following 
the detailed evolution of a system from general initial conditions, 
most analytic work has focused on studying the evolution of isolated 
spherical systems (Gunn \& Gott 1972).  The collapse of a spherical density
perturbation of a self-similar form, $\delta \propto r^{-m}$, in an
otherwise flat universe, yields the density profile $\rho \propto
r^{-3m/(1+m)}$ for $m\ge 2$, and $\rho\propto r^{-2}$ for $m<2$
(Filmore \& Goldreich 1984, Bertschinger 1985).  This implies that 
in highly non-linear systems, full information about the
initial distribution is preserved only for a special class of initial
conditions.  Strictly speaking, this result is valid only for the case
of purely radial collapse. In principle, non-radial motions can
prevent particles with large turnaround radii from sinking to the
inner regions of the collapse and forming an $r^{-2}$ profile.  If
particles were assigned angular momenta in a self-similar way, then
the density profile above is expected to be valid 
when $m<2$ as well (White \& Zaritsky 1992).

 These special spherical solutions were first
related to the formation of dark matter halos from initial gaussian
density fields by Hoffman \& Shaham (1985) (also see Hoffman 1988).
They noted that the mean shape of high peaks in gaussian fields is
$\delta \propto r^{-(n+3)}$ where $n$ is the index of the initial
power spectrum (Dekel 1981, Bardeen et al. 1986). Therefore, they
argued that setting $m=n+3$ makes the spherical solutions relevant in
the cosmological context. There are two caveats to this argument.
First, it assumes that the accreting matter is not clumpy.  In
contrast, in hierarchical scenarios for structure formation, halos
grow in mass by mergers with other, typically less massive, halos
(cf. Lacey \& Cole 1994; Lemson 1995; Syer \& White 1996).  
This means that the effects of dynamical friction
and tidal stripping may well be important in determining the final
structures of halos in hierarchical models.  Second, one is usually
interested in the density profiles of halos of a given mass today;
these may have had, as their seeds, initial density peaks of various
heights and shapes.  Moreover, strictly speaking, the Hoffman \&
Shaham (1985) solution applies only in regions outside the virial
radius of a dark matter halo.  In fact, Syer \& White (1996) argue
that the profile must scale as $r^{-\beta_{\rm SW}}$ with $\beta_{\rm
SW}={3m/(2+m)}$ in the inner regions of dark matter halos, and this
solution differs from the $r^{-\beta_{\rm HS}}$, with $\beta_{\rm
HS}={-3m/(1+m)}$, scaling proposed by Hoffmann \& Shaham (1985).

Since understanding the density profiles of halos is necessary to
relate structure formation theory to the observed structure of
galaxies, we aim here to develop a more detailed model of the process
of halo formation.  We will continue to assume spherical symmetry, and
will formulate a semi-analytic scheme for modelling the evolution and
the structure of halos more realistically.  The first goal of this
paper is to develop and test a method for tracing the mass of a given
halo back in time that is consistent with the hierarchical clustering
scenario.  We describe this method briefly in section 2.1 (details are
discussed in the Appendix).  Our second goal is to provide a simple
model, which includes the effects of dynamical friction and tidal
stripping, for the motion of satellites once they are inside a halo,
and to then use this model to estimate the halo structure that
results.  In section 2.2 we describe how to treat the motion of
satellites once they are inside the halo.  The equation of motion must
be solved numerically and, in section 3, we present the results of
applying this scheme to simulate the evolution of a variety of halos.
We study halos of various masses that form from initially scale free
gaussian fluctuation fields as well as fields with the CDM power
spectrum. In section 4 we conclude with a discussion of our results
and argue that the concentration scale $c$ in the NFW profile, if
fundamental, cannot directly be related to the formation time and must
reflect an interplay between the formation time and other---yet
unknown---physical effects, which determine the halo profile.

\section{The Model}
We will compute the evolved density profile of a halo which has mass
$M_0$ at the final redshift $z_0$ in two steps.  First, we devise a
scheme for tracing back, to some early time, the mass of the largest
progenitor subclump of the halo.  The rate of change of the mass of 
the most massive progenitor (MMP) gives the accretion rate of matter onto the already existing 
halo.  This rate is sensitive to the background cosmology, and to the 
shape of the initial power spectrum.  Our second step is to formulate 
a dynamical prescription for following the evolution of the 
accreted matter inside the halo.  

\subsection{Halo Mass Growth Rate}
At any given redshift $z$, define the current virial radius, $r_{\rm v}$, of 
a halo as the radius within which the average density is $178\rho_b$, 
where $\rho_b$ is the mean density in the universe at that time.  
At $z$, the current halo mass is defined as the mass within the 
current virial radius.  Ofcourse, the current virial radius and mass are 
smaller than their final values. Two key ingredients in our model are 
$(i)$ how the mass within the virial radius grows with time, and 
$(ii)$ how the accreting mass is distributed among satellites. 
To provide these two ingredients, one must trace the full merger 
history of a given halo back in time.  That is, we need to know 
how the mass $M_0$ was partitioned into progenitor halos at any 
earlier time $z_1>z_0$.  For Gaussian initial conditions, merger 
history trees of halos, each of which has the same mass $M_0$ at 
$z_0$, can differ substantially.  Therefore, we need to list all 
possible merger trees, and we need to compute the probability 
that each occurs.  Except for Poisson initial conditions 
(Sheth 1996), this probability has not been computed analytically. 
Hence, we adopt a short-cut.  We trace only the most massive 
progenitor  of the most massive progenitor, and so on, 
back in time.

We do this as follows.  We assume that at $z_1$, the average number of
progenitors of an $(M_0,z_0)$-halo that have mass between $M_1$ and
$(M_1+{\rm d}M_1)$ can be approximated by
\begin{eqnarray}
N(M_1,z_1|M_0,z_0)\,{\rm d}M_1 &=& \left(\frac{M_0}{M_1}\right)
\,\frac{1}{\sqrt{2\pi}}
\frac{\delta_{\rm c}(z_1-z_0)}{(S_1-S_0)^{3/2}}\,
\exp\left[-\frac{\delta_{\rm c}^2\,(z_1-z_0)^2}{ 2(S_1-S_0)}\right]
\left\vert\frac {{\rm d} S_1}{{\rm d} M_1}\right\vert \,{\rm d} M_1 
\nonumber \\
&\equiv& \left(\frac{M_0}{M_1}\right)\,f(M_1,z_1|M_0,z_0)\,{\rm d}M_1 
\label{ave-num} 
\end{eqnarray}
(Bower 1991, Lacey \& Cole 1993),
where $\sqrt{S_i}$ is the $rms$ density fluctuation in a Top-Hat
window function of radius $(3M_i/4\pi)^{1/3}$.  Lacey \& Cole (1994) 
show that this expression is in good agreement with what happens 
in numerical simulations (but see Tormen 1998).  The second 
equality defines $f(1|0)$, which is the fraction of the mass of 
$M_0$ that, at $z_1$, was in objects with mass $M_1$.  This means, 
of course, that the integral of $f(1|0)$ over the range 
$0\le M_1\le M_0$ equals unity.  However, as $(z_1-z_0)\rightarrow 0$, 
the integral of $N(M_1,z_1|M_0,z_0)$ over the range 
$(M_0/2)$ to $M_0$ also approaches unity.  Since an $M_0$ halo may 
have at most one progenitor with mass in this range, $N(1|0)$ can 
be interpreted as the probability that an $(M_0,z_0)$-halo had an 
$(M_1,z_1)$ progenitor subhalo (cf. Lacey \& Cole 1993), with 
$(M_1/M_0)$ restricted to the range between one half and unity, 
a short time earlier.  
Consequently, in the limit of small redshift intervals, we 
choose the mass of the MMP according to equation~(\ref{ave-num}).  
We then replace $M_0$ with the chosen value of $M_1$, and $z_0$ 
with $z_1>z_0$, and iterate until $M_1$, the mass of the MMP is 
as small as desired.  
In the Appendix we argue that this scheme provides MMP histories 
that are similar to those which occur in numerical simulations.  

\begin{figure}
\centering
\mbox{\psfig{figure=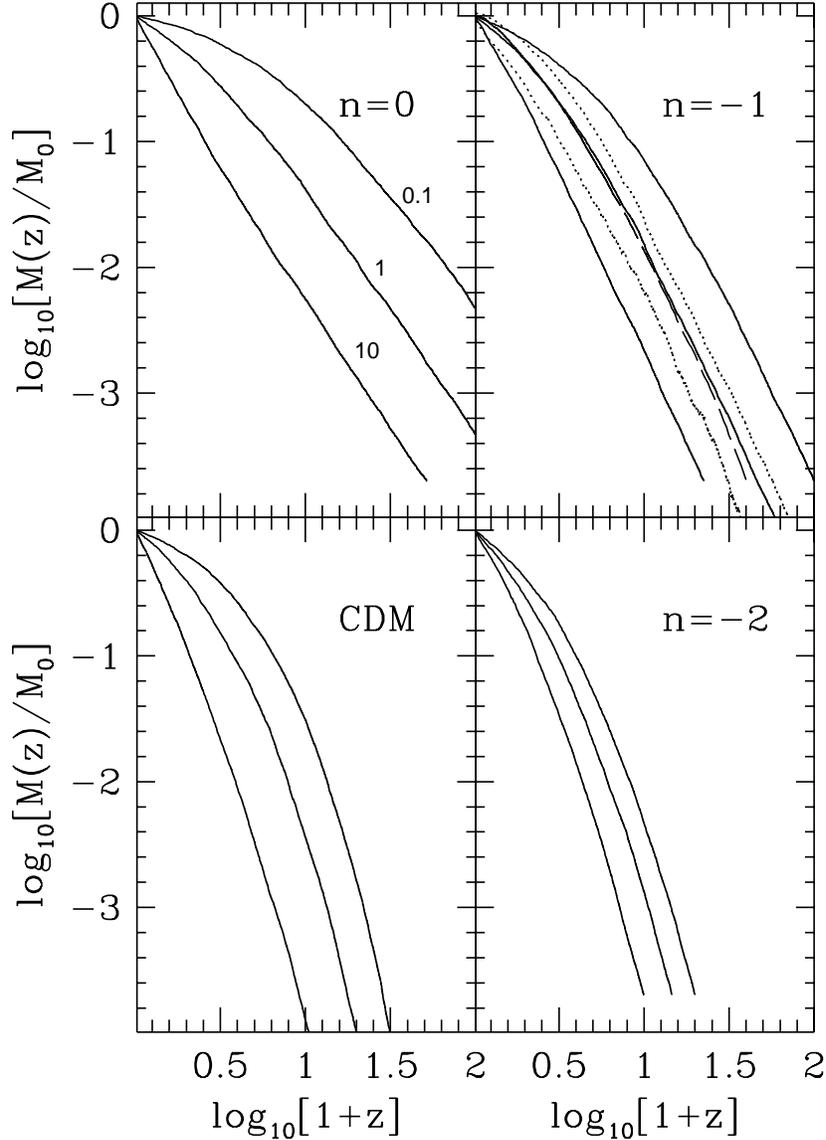,height=18.0cm}}
\caption{Curves of the average $M(z)/M_0$ for power law spectra with
$n=-2$, -1 and $0$, as indicated in the plot. The solid curves, in each
panel, correspond, from top to bottom, to $M/M_*=0.1$, 1 and 10
obtained by averaging in the mass direction.  The dotted lines in the middle
panel lines are $1\sigma$ deviations about the mean $M(z)$ for $1M_*$
halo. The dashed line in the middle panel is $M(z)$ for $1M_*$
obtained by averaging in the redshift direction.}
\end{figure}

In what follows, the ensemble average of the accretion rate will 
be useful.  Before computing it, the following question arises:
should we use curves of $M(z)$ by averaging the masses of all
realisations of merger histories in the mass direction given the
redshift, or should we use $M(z)$ obtained by averaging the
redshift at a given mass?  Fortunately,
the two ways of averaging
lead to almost identical curves of $M(z)$. 
Figure~1 shows curves of $M(z)$ for halos having $M_0/M_*=1,3$ and 
$10$, for scale free power spectra with $n=0,-1$ and $-2$. 
Also shown are results for halos having $M=10^{11}, 10^{13}$ and 
$10^{15}M_\odot$ in the standard CDM model, normalised such that 
the linear $rms$ value of density fluctuations in a spherical window 
of radius $8h^{-1}$Mpc is $\sigma_8=0.6$.  The solid lines 
show curves obtained by averaging the mass at a given redshift, 
and the dashed lines show curves obtained by averaging the redshift 
at a given mass.  The error bars are 1-$\sigma$ deviations from 
the mean $M(z)$.  The dashed and the solid lines corresponding to 
each case almost overlap.  This figure also shows the well known fact that, on 
average, lower mass halos formed at an earlier epoch.
 In what follows we have arbitrarily chosen 
to work with $M(z)$ obtained by averaging the merger histories in 
the mass direction. \footnote{
When the mass of the MMP is greater 
than half of $M_0$, then $M(z)$ can be computed directly 
from the formation time argument given in the Appendix.  
The mean value of the scaled variable there, $\omega_{\rm f}$, is 
\begin{displaymath}
\int_0^\infty \omega_{\rm f}\ p(\omega_{\rm f})\,{\rm d}\omega_{\rm f}
= \sqrt{2\over\pi}\left(\frac{2}{3} + \frac{1}{3f}\right),
\qquad (n=0)
\end{displaymath}
from which the mean value of $z$ given $M(z)/M_0$ is easily obtained.}

\subsection{Density profiles from stable clustering}

One way to convert the mass growth curves $M(z)$ to density 
profiles is as follows.  
We assume that the mass $M(z-{\rm d}z) - M(z)$ 
accreted during ${\rm d}z$ is accreted smoothly onto the MMP.  
This means that we are ignoring the fact that this mass may actually 
be divided among subclumps.  
Assume that a newly accreted patch of matter oscillates 
around the center of the halo with mean radius equal to the virial 
radius of the halo at the redshift at which it joined the halo. 
This is the {\it stable clustering} assumption, since it is equivalent 
to assuming that there is no net in-flow of matter in the virialised 
region.  It is motivated by the analytic solutions for collapse 
of density peaks with pure power law initial profiles 
(Filmore \& Goldreich 1984; Bertschinger 1985).  Recall that, 
in these solutions, the mean density within a radius $r$ of the 
halo is a multiple (here we use 178) of the background density 
of the universe at the time when the virial radius of the halo was $r$.  
Indeed, in this approximation, our only 
change to the Hoffman \& Shaham (1985) study is our method for 
computing the accretion rate. A similar analysis was also done by 
Avila-Reese, Firmani \& Hern\'andez (1997) who used the prescription
of  Lacey \& Cole (1993) and Eisenstein \& Loeb (1996) to compute $M(z)$.

\begin{figure}
\centering
\mbox{\psfig{figure=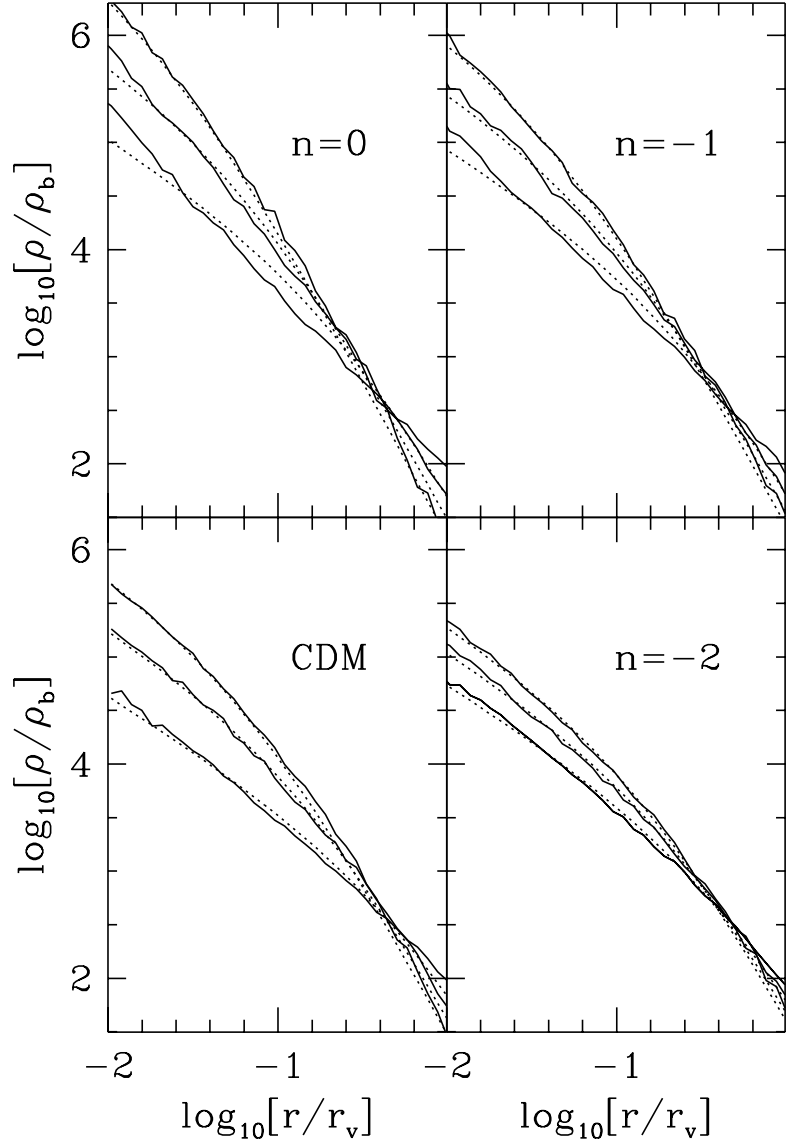,height=18.0cm}}
\caption{ Density profiles obtained from curves of $M(z)$ 
under the assumption of stable clustering. The dotted 
curves are the NFW best fits
to the density profiles.}
\end{figure}

\begin{figure}
\centering
\mbox{\psfig{figure=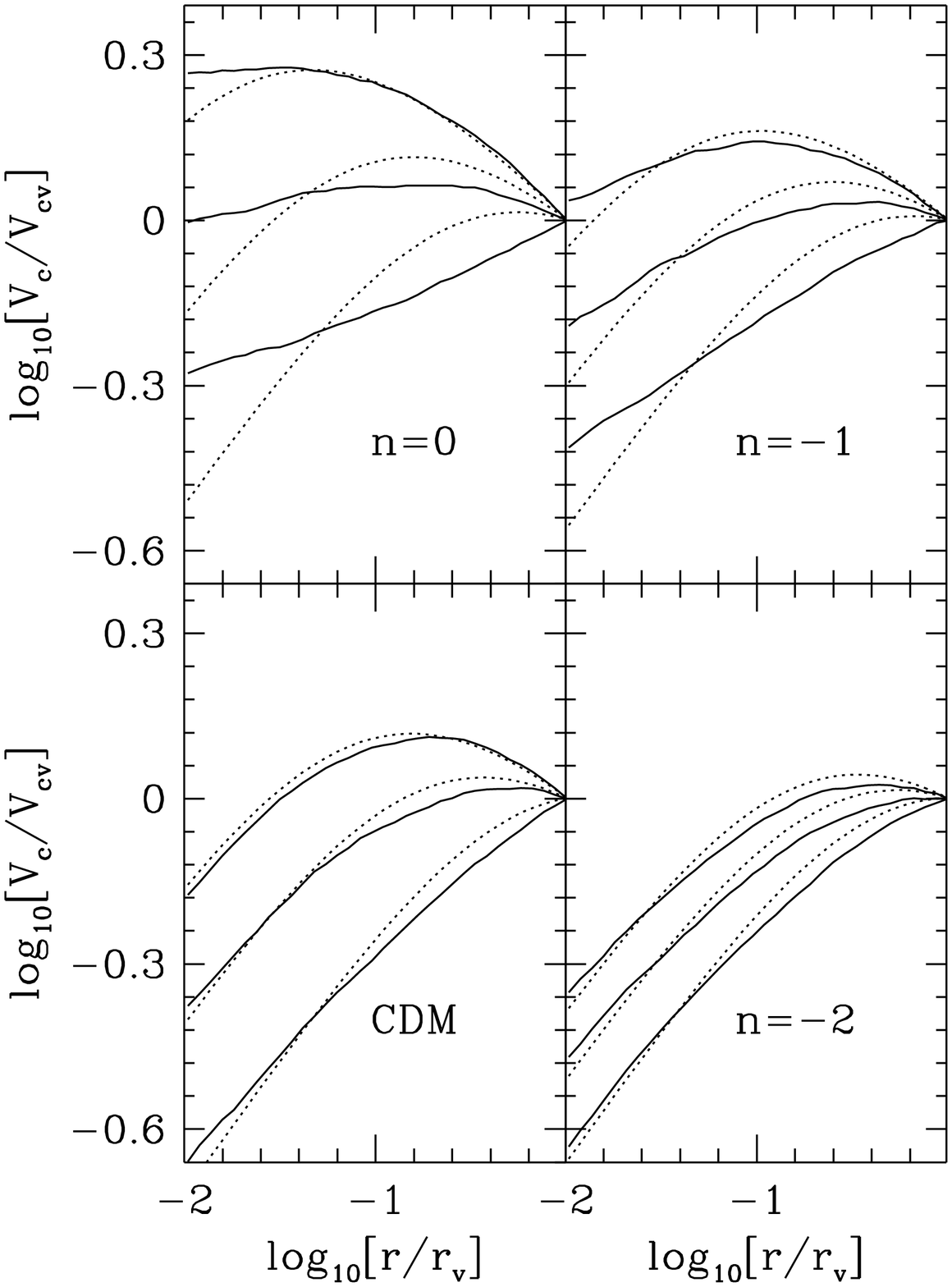,height=18.0cm}}
\caption{ The same as the previous figure but showing rotation curves
instead of density profiles.}
\end{figure}

Ideally, one would like to use stable clustering to compute final
profiles for each individual merger history and then average the
result over many realisations.  In practice this is time consuming.
We show below (Figure~4) that the evolved density profile computed
from the ensemble averaged $M(z)$ is a good approximation to the
ensemble average of the individual evolved profiles.  
To do so, we have used the stable clustering assumption to compute 
profiles of 10 random realisations of the merger tree of an $M_*$ 
halo for a scale free spectrum with $n=-1$. The average of these 
profiles is the dashed curve in figure~4. The profiles from six 
individual merger histories are shown by the light solid curves. 
The thick solid line is the profile obtained from the average
$M(z)$ shown in figure~1. We conclude from figure 4 that, at least 
under the assumption of stable clustering, working with the average 
$M(z)$ does not introduce systematic biases.  

The density profiles generated in this way from the ensemble averaged 
$M(z)$ curves of Fig.~1 are shown in Fig.~2.  The dotted lines are
fits of the form NFW form given in (\ref{nfw}).  
Figure~3 shows the associated circular velocity profiles:  
$V^2_{\rm c}(r) \propto GM(r)/r$.  
Except when $n=0$, the individual density and circular velocity 
profiles are reasonably well described by the NFW parametric form.  
Notice that less massive halos are more centrally concentrated. 
The trend of steeper profiles for larger values of the spectral 
index, $n$, is also seen in the simulations (e.g. Cole \& Lacey 1996). 
There is, however, a fundamental difference between the profiles 
obtained here and those in the simulations.
For example, in the CDM N-body simulations, profiles of halos with 
masses $\sim10^{11}M_\odot--\sim 10^{15}M_\odot$ almost coincide over 
the range $0.1>r/r_{\rm v}<1$ (cf. figure~4 in NFW 1995). 
On the other hand, the corresponding profiles in figure~2 
substantially deviate from each other for the same range.  
Thus, the concentration parameter $c$ depends differently on mass 
in our models than it does in the simulations.  
NFW argue that $c$ is determined by the formation time of halos.  
Our detailed models of the accretion history incorporate this 
formation time explicitly, yet, they are unable to reproduce the 
trends of $c$ with mass that are seen in simulations.  This suggests 
that, if simulated halos are indeed well described by the one
parameter $c$, then this parameter must depend on more than just 
the formation time; $c$ must depend on some additional physics.  

The stable clustering assumption is not expected to hold in realistic
collapse configurations. This is because the accreting matter is, 
in fact, distributed into bound clumps (satellites).  These will 
suffer from dynamical friction as they fall into the halo.  This 
causes a transfer of energy from the satellites to the halo which 
is neglected in the stable clustering collapse considered above.  
Moreover, the tidal field of the halo continuously prunes 
infalling satellites, and this is also ignored in the smooth 
collapse scenario. In the next section we describe how to 
compute evolved density profiles which include these effects. 

\begin{figure}
\centering
\mbox{\psfig{figure=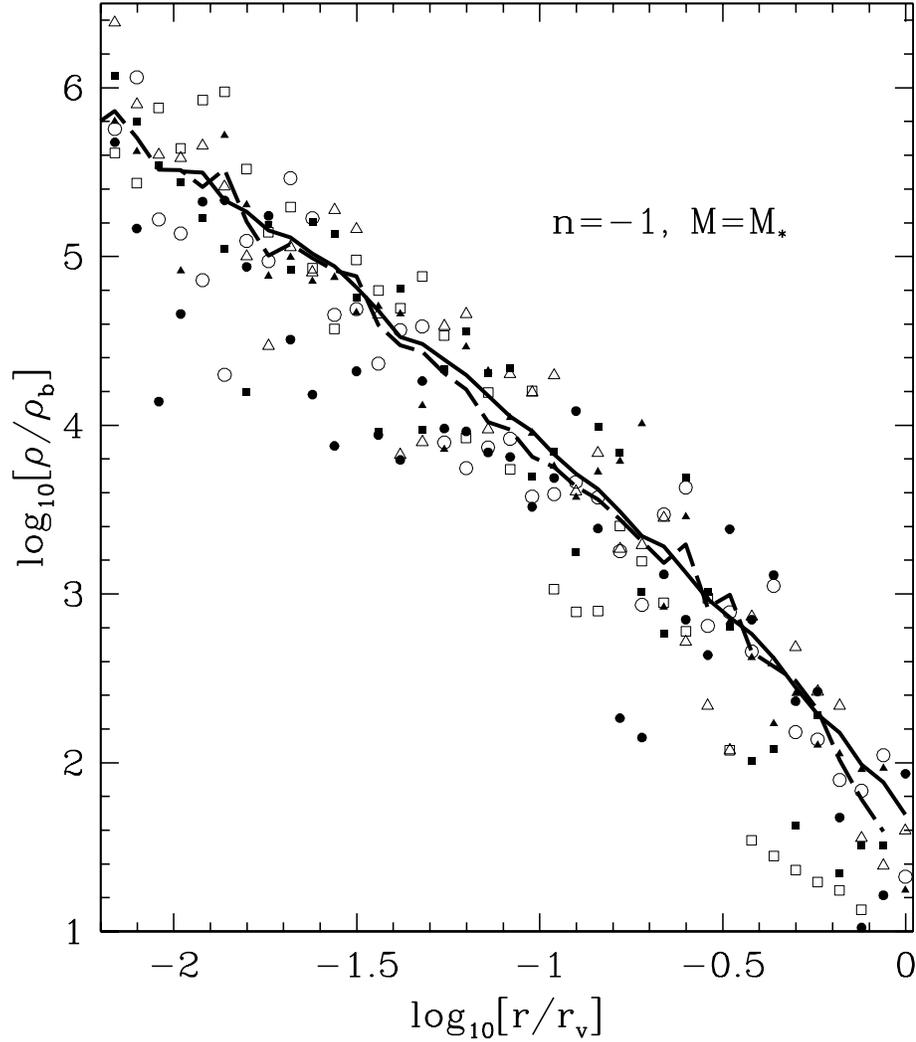,height=18.0cm}}
\caption{ Density profiles from 6 random realisations (points) of $M(z)$ 
for $n=-1$ and $1M_*$ halo. 
Plotted also is the average profile (heavy dashed) of 10 random
realisations and the profile (heavy solid) corresponding to the average
$M(z)$.
The profiles were computed under the  stable clustering assumption.}
\end{figure}

\subsection{Beyond stable clustering:  
dynamical friction, tidal stripping and halo heating}

In this section we formulate a dynamical model of the motion of 
satellites inside halos.
The treatment here does not require any assumptions
about the matter accretion rate, provided major mergers
do not occur. Indeed  mergers of two halos of similar mass 
invalidate the  dynamical prescription presented here.  

We will write the equations of motion of satellites inside a halo
under the approximation that the halo is spherical.  
Then, the gravitational field of the halo is that of a spherical 
mass distribution. In addition to the gravitational field of the 
halo, satellites suffer from a dynamical friction drag force 
(Chandrasekhar 1943).  Therefore, the equations governing the 
motion of a satellite within a halo are,
\begin{equation}
{{\rm d}^2 {\bf r}\over {\rm d}t^2}=
- \omega_{\rm df}(r,t) {{\rm d}{\bf r}\over {\rm d}t} 
- M_{\rm h}(r,t)\frac{\bf r}{r^3} ,
\label{eq-motion}
\end{equation}
where $M_{\rm h}(r,t)$ is the mass of the halo within radius $r$ 
at time $t$ and $\omega_{\rm df}$ is the inverse of the dynamical 
friction time.\footnote{We work in units in which $G=1$.}  
Ostriker \& Turner (1979) derived a similar expression.  
For a satellite of mass $M_{\rm s}$ moving in a medium of density, 
$\rho$, and velocity dispersion, $\sigma$, the quantity 
$\omega_{\rm df}$, can be approximated by 
\begin{equation}
\omega_{\rm df}(r,t)=\cN\,M_{\rm s}\,{{\rho(r,t)}\over
{\sigma^3(r,t)}} \label{tdf}
\end{equation}
(Binney \& Tremaine 1987), where $\cN$ is a numerical factor. 
Here $\sigma(r,t)$ and $\rho(r,t)$ represent the average velocity 
dispersion and density within $r$.

The gravitational field of the halo prunes the satellites as they 
sink to the center of the halo. This causes a gradual decrease in the 
mass  of the satellite, $M_{\rm s}$, which should be taken into account 
in the equations of motion ~(\ref{eq-motion}).
We assume that a satellite at a distance $r$ from the center of 
the halo retains only matter which lies inside a
certain radius, the tidal radius $r_{\rm t}$, from its center. 
This tidal radius is approximately that radius at which the 
gravitational field of the satellite equals the change in the 
gravitational field of the main halo times $r_{\rm t}$.  That is, 
\begin{equation}
{{\partial} \over \partial r}\!
 \left({{M_{\rm h}}\over  {r^2}}\right) 
r_{\rm t} = {{M_{\rm s}}\over {r_{\rm t}^2}} ,
\end{equation}
which, to first order in $r_{\rm t}/r$, reduces to
\begin{equation}
{{M_{\rm s}(r_{\rm t})} \over r^3_{\rm t}} 
= {{M_{\rm h}(r)}\over r^3} .
\label{eq-tidal}
\end{equation}
Thus, the average density of the satellite within $r_{\rm t}$ 
equals the average density of the halo within $r$.  Given a guess 
about the density profile of the satellite when it crosses the 
virial radius of the halo, and assuming that the profile within 
$r_{\rm t}$ does not change with time, 
the relation ~(\ref{eq-tidal}) determines the mass of the satellite 
as a function of $r$.

Here we assume that satellites are spherical and have $1/r^2$ 
density profiles, i.e., $M_{\rm s}(r_{\rm t})\propto r_{\rm t}$.  
In this case, as a satellite of mass $M_{\rm sv}$ passes 
through a halo of density $\rho_{\rm v}$, it loses mass according to
\begin{equation}
M_{\rm s}(r) = 
\left({{\rho_{\rm v}} \over {\rho_{\rm h}(r)}}\right)^{1/2} 
M_{\rm sv},  \label{satmass}
\end{equation}
where $M_{\rm s}(r)$  is the mass  of  the satellites
mass when it is at radius $r$ and $\rho_{\rm h}(r)$ is the average
density of the halo within that radius.
Substituting this in~(\ref{tdf}), and using the fact that 
$M_{\rm h}(r)/r\propto \sigma^2(r)$, we find that 
\begin{equation}
\omega_{\rm df} = {\alpha} \left(\frac{\rho_{\rm v}}{3\pi}\right)^{1/2} 
{{M_{\rm hv}} \over {M_{\rm h}(r)}} ,
\label{tdfm}
\end{equation}
where  $\alpha=\sqrt{3\pi}{\cal N} M_{\rm sv}/M_{\rm hv}$, and 
$M_{\rm hv}$ is the mass of the halo when the satellite crossed 
its virial radius. The factor $\alpha$ is scaled so that it is the 
ratio of the circular orbit time to the dynamical friction time 
at the time when the satellite joined the halo.

Matter stripped from satellites is said to be `evaporated'.  
Initially, evaporated matter is assigned the velocity and position 
of the satellite from which it was stripped.  Thereafter, it moves 
under the influence of the gravitational field of the halo 
free of dynamical friction.  Of course, the satellite, now 
with reduced mass, continues to suffer from dynamical friction.  

Next, we need to determine the fate of the energy that 
is lost by satellites as a result of the dynamical friction 
term in the equation of motion~(\ref{eq-motion}).  
A satellite loses energy at the rate  
\begin{equation}
{{\rm d} (E/M_{\rm s})\over  {\rm d}t}=-\,\omega_{\rm df} 
\left({{\rm d} {\bf r}\over {\rm d}t}\right)^2   
\label{eqener}
\end{equation}
(cf. Landau \& Lifshitz 1960).  
It is unclear how this lost energy is redistributed within the halo. 
Since the bulk of the halo's matter contributes to the 
dynamical friction drag force which is exerted on the satellites,
it seems reasonable to assume that the lost energy is distributed
equally among the halo matter.  That is, the energy deposited 
to a patch of evaporated matter is assumed to be proportional to its 
mass. 

\subsubsection{Numerical Scheme}
The equations of motion above can be solved numerically, 
provided that we make a number of additional assumptions which 
we describe below.  

Our numerical scheme simulates a halo made of particles of identical
mass. Some of these particles are assigned to satellites and suffer
from dynamical friction; the other particles move under the influence
of the gravitational field of the halo only and are called halo
particles.  The code treats these two types of particles differently.
Since a satellite particle may be tidally stripped from its parent, we
need some way of deciding when this happens; once stripped the
particle becomes a halo particle, and can never be captured by
satellites.  Dynamical friction provides a mechanism for the transfer
of energy from satellites to the halo; the code allows for this by
assuming that only halo particles can absorb this energy.  We describe
how both these effects are included in more detail below.
  
We follow the evolution of a halo starting at an early time at 
which the halo mass was a tiny fraction of its final mass. 
Given the growth of the mass of the halo with time, we can compute 
the time at which each particle will cross the virial radius of the 
halo. At any given time, we simulate only the trajectories of 
those particles which are within the current virial radius. 
We assume that, when first accreted, particles move on circular 
orbits.

Since we are only simulating particles within the current virial
radius, some of these particles may have already been stripped 
from their parent satellites. 
Here we assume that about half the particles joining the halo in
any given time step are associated with satellites and the other half
are halo particles.   
We need to specify the parameter $\alpha$ for all 
those particles which belong to satellites.
Notice that equations~(\ref{eq-motion}) and~(\ref{tdfm}) 
involve the mass of a satellite only through the parameter 
$\alpha$ which determines the dynamical friction time scale.  
This simplifies the problem considerably.  
In the numerical scheme, each satellite's particle is assigned a 
value of $\alpha$ drawn from a uniform distribution with a 
minimum value of zero and some maximum value, $\alpha_{\rm max}$, 
which determines the total energy lost by the satellites.

This choice implies that, at any time, the average satellite mass 
is proportional to the mass of the MMP. 
This is in the spirit of hierarchical clustering
and is motivated by results from N-body simulations (Tormen 1997).  
Now we need to decide which particles are stripped, and when.
At any given time, each satellite particle is assigned a probability
of remaining bound to its parent.    
According to equation~(\ref{satmass})  this probability
is proportional to the ratio of the average density within 
the particle's radius at the current, to that at the previous 
time step.

Finally, in order to avoid numerical instabilities, the gravitational
force acting on a particle at radius $r_i$ is softened according to
\begin{equation}
{\bf F}_{{\rm g}i}=-\frac{m N_i{\bf r }_i}{(r_i^2+\epsilon^2)^{3/2}}
\end{equation}
where $m $ is the particle's mass, $N_i$ is the number of particles
with $r<r_i$ and $\epsilon$ is the force softening parameter.  
For a closed system of particles without dynamical friction this
force law conserves the total ``energy'' 
\begin{equation}
E_{tot} = m\sum_i\left[\frac{{\dot {\bf r}_i}^2}{2}
-\frac {mN_i}{\left(r_i^2+\epsilon^2\right)^{1/2}}\right] 
\end{equation}
(White 1983).
The factor $\omega_{\rm df}$, which determines the amplitude of the
dynamical friction drag force, is evaluated according to
\begin{equation}
{\omega}_{{\rm df}\ i} = 
\alpha_i \,\rho^{1/2}_{{\rm v}i}\,\frac {M_{{\rm hv}i}}{mN_i} ,
\end{equation}
where  $M_{{\rm hv}i}$  and $\rho_{{\rm v}i}$ are the mass and the 
average density of the halo within the virial radius at the time 
the satellite was accreted onto the halo.  The sorting routine 
INDEXX in Press et. al. (1992) reduces the evaluation 
of the gravitational and dynamical friction forces to an 
$N\log_2 N$ process, where $N$ is the total number of particles in
the halo at any given time step.  Once the forces have been computed,
a leapfrog time integration scheme is used to move particles from the
current time step to the next.  In each time step the energy lost by
satellite particles is computed using ~(\ref{eqener}).  
The satellites lost energy is distributed equally 
among the halo particles in the form of kinetic energy. 
Namely, halo particles absorb the energy lost by satellites by 
receiving ``kicks'' in random directions. 

Increasing $\alpha_{\rm max}$ increases the total amount of 
energy lost by the satellites. Unfortunately, the code is unstable for
values of $\alpha_{\rm max}$ which lead to energy loss exceeding a
significant fraction of the absolute value of the total energy.
We will see that even a small amount of heating is sufficient to
significantly change the halo density profile in the inner region
relative to the stable clustering profiles.

\section{Results}

We ran simulations with 80000 particles in the final halo and with
30000 equally-spaced time steps. In all the runs we set the softening 
parameter $\epsilon=0.0005\,r_{\rm v}$, where $r_{\rm v}$ is the 
virial radius of the halo at the final time.  For convenience we 
adjust the numerical factors in the equation such that $\alpha$ is 
the ratio of the dynamical friction time to the circular orbit time 
when the satellite just joined the halo.
Following each time step, the satellites' lost energy is reassigned
to particles in the halo. 

It is interesting to inspect the effects of dynamical friction and
tidal stripping. In figure 5, we plot trajectories of satellites
moving in a halo with a density profile given by that of a $10^{15}M_\odot$
 CDM halo as obtained assuming
stable clustering and shown in Fig.~2.  The
satellites were initially placed on circular orbits in the XY plane.
The orbits were integrated with $\alpha=3$ and $1/2$ respectively.
For comparison a trajectory with no tidal stripping is also shown.
\begin{figure}
\centering
\mbox{\psfig{figure=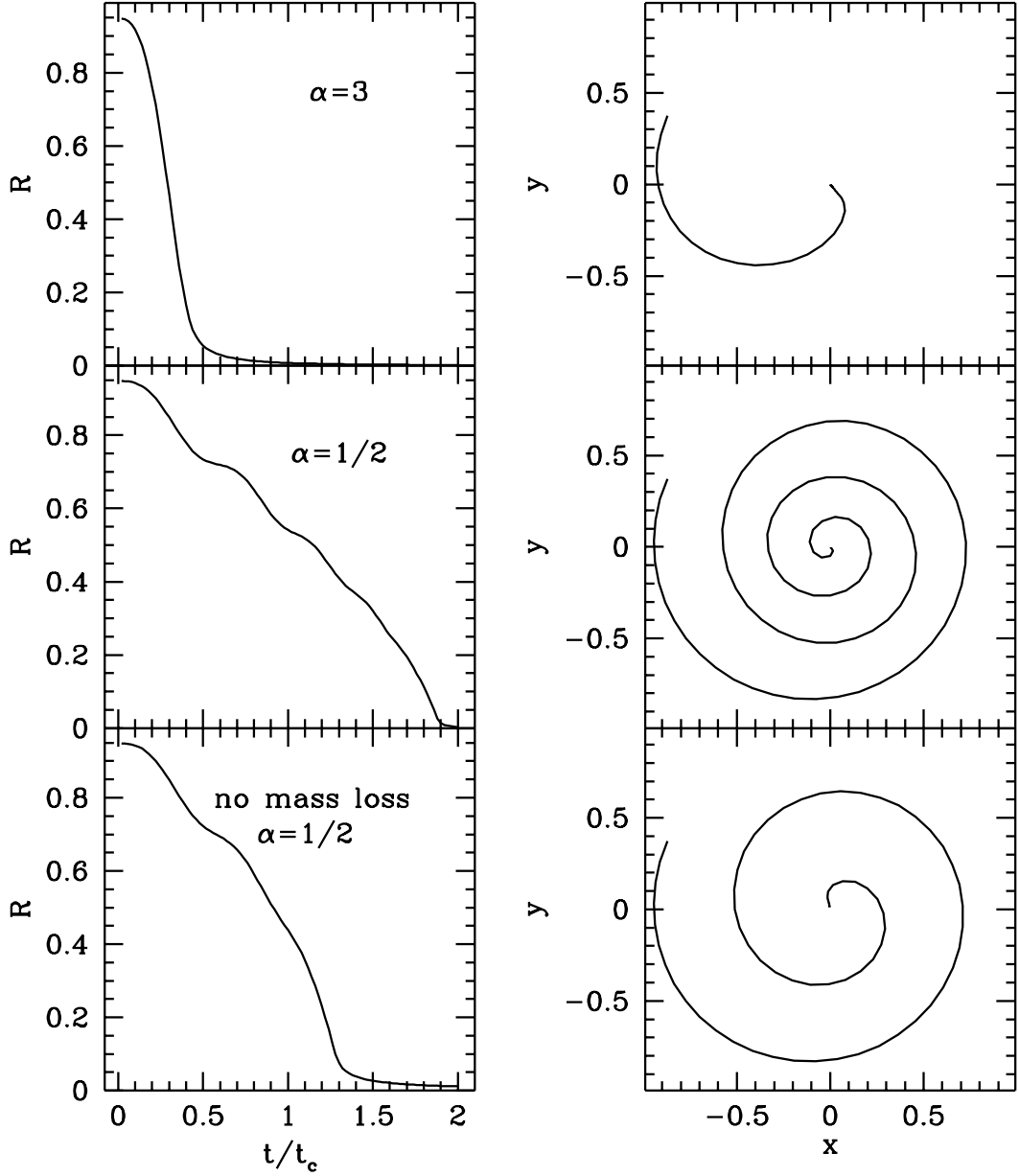,height=18.0cm}}
\caption{Orbits of satellites with various values of $\alpha$.
The stable clustering density profile of a CDM $10^15M_\odot$ 
was used to integrate the equations. The bottom panel shows an orbit if
satellites did not lose mass as a result of tidal stripping.}
\end{figure}

Fig.~6 shows density profiles of halos at the final time.  Results
are shown for halos with $M/M_*=0.1,1$ and 10 for the scale free power
spectra and $M=10^{11},10^{13}$ and $10^{15}M_\odot$ for the standard
CDM model.  The dotted lines show the best fit NFW profiles.  The
curves were computed assuming the satellites' lost energy was
redistributed in the form of kinetic energy.  The profiles in Fig.~6
were computed assuming $\alpha_{\rm max}=1/2$. For the CDM model, the
fraction of the total energy lost by satellites for this 
value of $\alpha_{\rm max}$ was $0.23$, $0.11$ and
$0.04$, for halos of mass $10^{11}$, $10^{13}$ and $10^{15}M_\odot$,
respectively.  Fig.~7 shows the circular velocity profiles
corresponding to the density profiles shown in Fig.~6.

Relative to the profiles obtained assuming stable clustering, 
these density profiles are steeper, at least in the inner regions.  
This is because the halo is heated by the infalling satellites. 
Recall that, in our prescription, energy lost by the satellites 
produces random motions among the halo particles. 
If the density profile is shallower that $r^{-2}$, then the 
number of particles crossing a radius $r$ inward, as a result of 
the added random motions, is larger than the number of particles 
crossing outward.  
We have tried various other heating recipes such as redistributing  
only half the lost energy in the form of kinetic energy while the other
half in the form of potential energy (by appropriately rescaling the 
radii of the halo particles).
The final profiles in all the heating recipes we adopted 
were similar but not identical; they all were steeper than the 
corresponding profiles obtained from stable clustering.

Although the density profiles for $n>-2$, are 
not very well described by NFW fits, the density profiles for 
$n=-2$ and CDM are.  The corresponding circular velocity profiles, 
plotted in Fig.~7, show significant deviations from the NFW fits. 
The slopes of the density profiles in the inner regions 
$r<0.1r_{\rm v}$ are listed in table 1. The top and bottom numbers
in each row correspond to the profiles obtained from stable 
clustering (Fig.~2) and from the profiles in Fig.~6, 
respectively.  It is interesting that the slope for the halo of 
mass $10^{15}M_\odot$ in the CDM model agrees with the result of 
the high resolution N-body simulation of Moore et. al. (1997).

\begin{figure}
\centering
\mbox{\psfig{figure=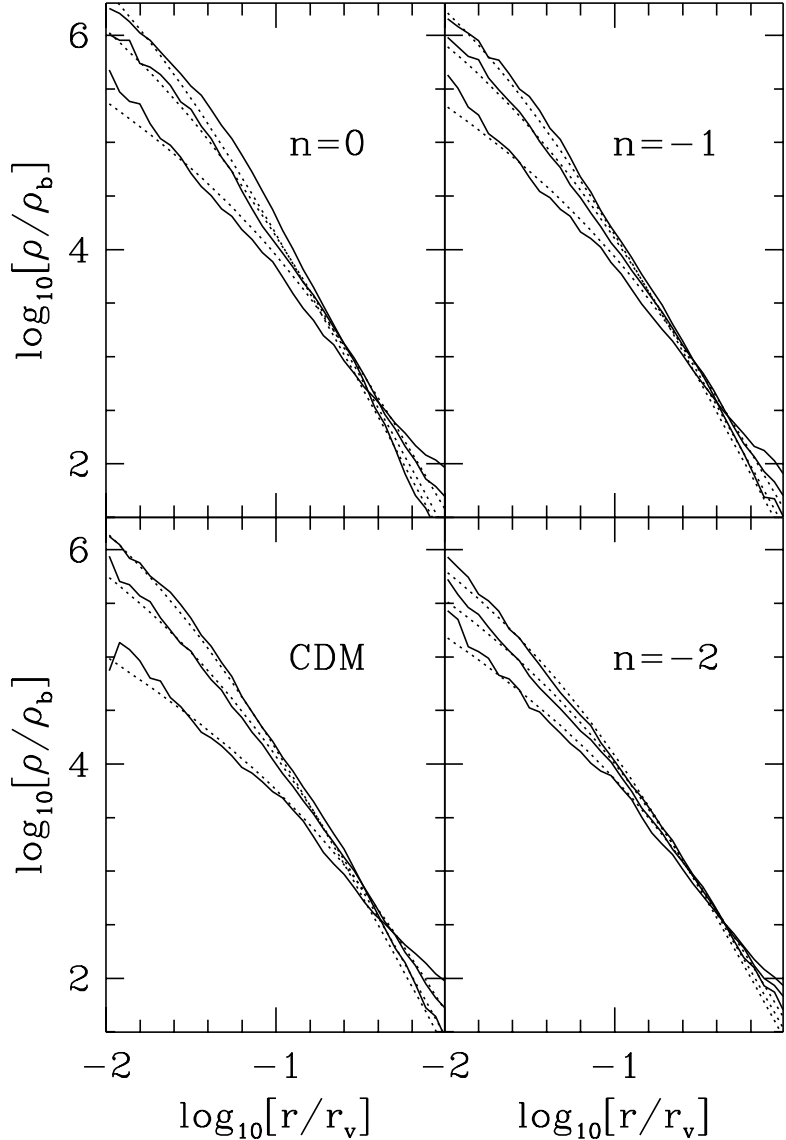,height=18.0cm}}
\caption{ Density profiles  obtained from the dynamical model using 
curves of $M(z)$ from Fig.~2.  
The dotted curves show the best NFW fits.}
\end{figure}

\begin{figure}
\centering
\mbox{\psfig{figure=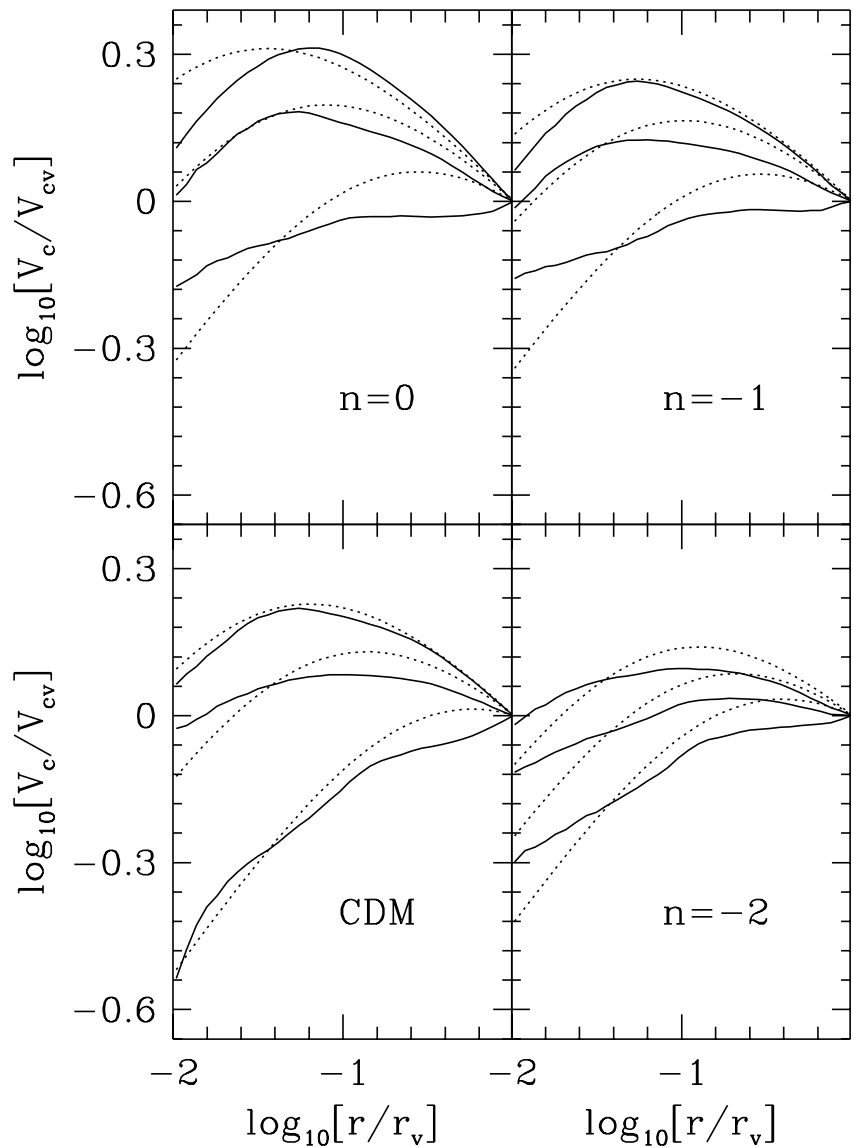,height=18.0cm}}
\caption{ The same as the previous figure but showing circular
velocity  profiles.}
\end{figure}

\begin{table}
\caption{Slopes of density profiles in the region $10^{-2}<r/r_{\rm v}
<10^{-1}$. For power law power spectra, $M_1, M_2$ and $M_3$
correspond to halos of masses $0.1,1$ and $10$ in units of $M_*$.  
For CDM, they correspond to $10^{11}$, $10^{13}$ and $10^{15}M_\odot$. 
Top and bottom values in each mass row are for the stable clustering 
and the semi-analytic dynamical model, respectively.}
\begin{tabular}{|c|c|c|c|c|c|} \hline
 \multicolumn{1}{|c|}{} & \multicolumn{1}{|c|}{$n=-2$}
&\multicolumn{1}{|c|}{$n=-1$} &\multicolumn{1}{c}{$n=0$} 
&\multicolumn{1}{c}{CDM} \\ \hline
$M_1$ & -1.45 & -1.84 & -2.12& -1.59\\ 
      & -1.94 & -2.04 & -1.94& -2.01\\ \hline
$M_2$ & -1.38 & -1.65 & -1.90& -1.41\\ 
      & -1.76 & -2.03 & -2.05& -1.91\\ \hline
$M_3$ & -1.26 & -1.54 & -1.76& -1.25\\ 
      & -1.60 & -1.77 & -1.84& -1.45\\ \hline
\end{tabular}
\end{table}

\section{Summary and conclusions}

Assuming gaussian initial conditions,
we addressed two related issues. These were: $a)$ how to
generate random realisations of the growth history 
of a halo whose mass at the present time is given, 
and $b)$ how the density profile of a halo can be related 
to its merging history.

The way we generate random realisations of halo histories is, in
principle, different from that previously presented in the literature.
Lacey \& Cole (1993) and Eisenstein and Loeb (1996) choose to work
with the function $f(M_1,z_1|M_0,z_0)$ which is the probability that a
mass element of halo $M_0$ at $z_0$ is incorporated in a halo of mass
$M_1$ at $z_1>z_0$. Instead, we work with
$N(M_1,z_1|M_0,z_0)=(M_0/M_1)f(M_1,z_1|M_0,z_0)$, which can be
interpreted as the average number of progenitors of $M_0$ that have
mass $(M_1)$ at $z_1$.  We have checked that these previous methods 
lead to results that are similar to ours, and we show in the 
Appendix why this is so.

Given the mass growth rate we presented results using two models for
computing the evolved density profiles.  The first was based on the
stable clustering assumption in which a mass element is effectively
assumed to remain at the virial radius when it joined the halo. 
The second invoked a more detailed description of the motion of 
satellites once they have crossed the virial radius of the halo, 
incorporating tidal stripping, dynamical friction and heating of 
halo particles by infalling satellites.  
We found that including these effects leads to steeper
profiles in the inner regions than in the stable clustering case.  For
initial power-spectra $P(k)\propto k^n$, we find that the slope in the
inner regions depends on $n$, and also on the final mass.  Table~1
shows that the inner profile is $\rho\propto r^{-\beta}$, with
$\beta_{\rm SW}\le\beta\le\beta_{\rm HS}$.  For a given power
spectrum, the dependence of density profiles on the halo mass in our
models is stronger than that seen in the simulations (e.g. NFW). In
particular, for the CDM power spectrum, the relative difference
between the profiles of halos of masses $10^{11} M_\odot$ and $10^{15}
M_\odot$ is about an order of magnitude at $r=0.1r_{\rm v}$. 
On the other hand, the corresponding profiles in the simulations 
almost coincide for $0.1<r/r_{\rm v}<1$.  This suggests that the NFW 
argument that the concentration parameter is entirely determined by 
the formation time is inadequate.  It is likely that effects, in 
addition to those included in our models, must play a significant 
role in determining halo density profiles.  
For example, in their study of fully three-dimensional collapse, 
Huss et al. (1998) suggest that a bar-related instability may 
play a major role in determining the final profile.  

Using our simplified models, we hoped to shed some light on the
current---arguably contradictory---results of various recent N-body
simulations (Navarro Frenk \& White 1997, Kravtsov et. al. 1997, Moore
et. al. 1997) of halo density profiles.  Based on these models,
the general conclusion is that a one parameter form, such as the fit
proposed by Navarro, Frenk \& White (1997), is inadequate to describe
the profiles for all radii and all power spectra.  

\section*{Acknowledgments}
We thank Simon White, David Syer and Gerard Lemson for many useful 
discussions. 

\protect{}

\appendix
\section{The algorithm}
This Appendix describes our MMP generating algorithm in detail.  
First we compare our algorithm with others that have been proposed 
in the literature.  Then we show that, in at least one respect, 
it is able to reproduce results measured in numerical simulations 
of hierarchical clustering.  This motivates further study:  
we show that, for our algorithm, the distribution of formation 
times and the distribution of formation masses can be computed 
analytically.  

\subsection{Comparison with previous work}
Recall that the MMP is followed through a 
succession of small timesteps.  At each step, the mass decreases 
according to equation~(\ref{ave-num}).  This means that the mass of 
the MMP is never smaller than half the mass of the parent halo at 
the previous time step.  It is hard to justify this assumption on 
general theoretical grounds (but see Sheth \& Pitman 1997 for a 
self-consistent, binary split model of the merger tree described 
by Sheth 1996).  
Nevertheless, this assumption is in agreement with results from 
N-body simulations;  Fig.~5 in Tormen et al. (1997) shows that 
the masses of the MMPs in their simulations do not decrease by 
more than a factor of two at each small time step.  

Our algorithm differs slightly from an MMP generating algorithm that
was suggested by Lacey \& Cole (1993), and used by Eisenstein \& Loeb
(1996).  They split the MMP according to $f(1|0)=(M_1/M_0)\,N(1|0)$, 
with no restriction on $M_1/M_0$ [recall that the integral of 
$f(1|0)$ over the range $0\le M_1/M_0\le 1$ equals unity as 
required], but they then choose the larger of the two pieces.  
This means that their MMP has mass $M_1$ with probability 
$(M_1/M_0)\,N(M_1|M_0)+(M_0-M_1)/M_0\,N(M_0-M_1|M_0)$, and 
$M_1\ge M_0/2$ always. In the limit $(z_1-z_0)\rightarrow 0$,
$N(M_1|M_0)\approx N(M_0-M_1|M_0)$, for a large range of $M_1$
(cf. Fig.~7 in Sheth \& Pitman 1997), so this is similar to using
$N(1|0)$ for the split rule.  The approximation of equality is good
for white-noise initial conditions, but it becomes worse for power
spectra with $n<0$.  

\begin{figure}
\centering
\mbox{\psfig{figure=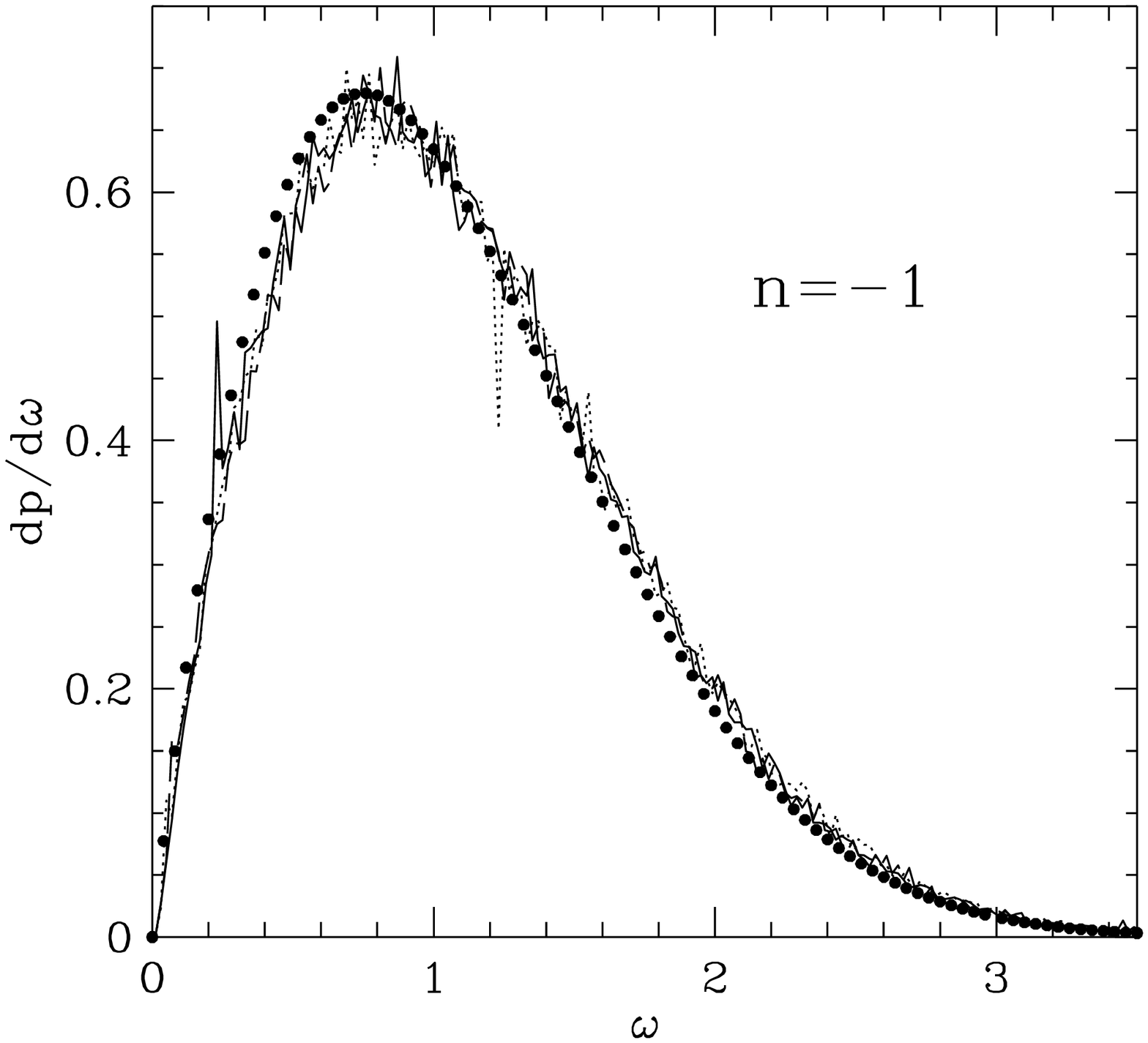,height=15.0cm}}
\caption{The distribution of the formation time of halos, 
expressed in terms of the variable 
$\omega = z\,(M_*/M)^{1/3}/\sqrt{(2^{2/3}-1) }$ for a power-law 
spectrum with slope $n=-1$. The  dotted, solid and dashed curves
correspond to $M/M_*=0.1$, 1 and 10, respectively.  
The points show equation~(2.33) of Lacey \& Cole (1993) for 
$n=0$.}
\label{zform}
\end{figure}

Let $t_{\rm f}$ denote the formation time, defined as the time when 
the mass $m$ of the MMP is a fraction $f$ of the halo at the present 
time.  Provided $f=m/M_0\ge 0.5$, this distribution can also be 
computed using the arguments presented in section~2.5.2 of Lacey \& 
Cole (1993).  For white noise initial conditions (initial 
power spectra with slope $n=0$), this distribution can be 
computed analytically.  If 
\begin{displaymath}
\omega_{\rm f} \equiv
\delta_{\rm c0}(z_1-z_0)/\sqrt{\sigma^2_m-\sigma^2_{M_0}},
\end{displaymath} 
then, provided $f\ge 0.5$, 
\begin{equation}
p(\omega_{\rm f}) = 
\left(\frac{1}{f}-1\right) 2\,\omega_{\rm f}\,
{\rm erfc}\left(\frac{\omega_{\rm f}}{\sqrt{2}}\right) 
+ \left(2-\frac{1}{f}\right) 
\sqrt{\frac{2}{\pi}}\,\exp\left(-\frac{\omega_{\rm f}^2}{2}\right),
\qquad\qquad(n=0).
\label{form}
\end{equation}
When $f=0.5$ this reduces to equation~(2.33) of Lacey \& Cole (1993).
For more general initial conditions the answer must be computed 
numerically.  For all $f\ge 0.5$, the distribution for other values 
of $n$ is very similar to that for $n=0$ 
(Fig.~7 of Lacey \& Cole 1993 shows this for $f=0.5$).  
At least when $f=0.5$, this distribution for the formation times 
fits the distribution measured in N-body simulations well 
(Lacey \& Cole 1994).  Since it is trivial to compute this
distribution from our Monte--Carlo MMP generating algorithm 
(simply follow the MMP till its mass drops to $f$ times the initial 
mass), it is important to check that our MMP generating algorithm 
is able to reproduce this distribution.  

As a test of our MMP history scheme, Fig.~\ref{zform} compares the 
distribution of formation times $p(t_{\rm f})$ obtained from our 
Monte--Carlo realizations with the analytic prediction given
by~(\ref{form}).  The distributions are plotted in terms of the 
scaled variable $\omega_{\rm f}$, and we have set $f=0.5$.  Thus, 
$(t_0/t_{\rm f})^{2/3}=1+\omega\sqrt{2^\alpha-1}(M_0/M_*)^{-\alpha/2}$ 
where $\alpha=(n+3)/3$ and $n$ is the slope of the initial 
power spectrum.  The figure shows results for $n=-1$, although 
our Monte--Carlo distributions for other values of $n$ are also 
in good agreement with the analytic prediction, as are the 
distributions for other values of $f\ge 0.5$.  The scatter of 
the Monte--Carlo distributions about the analytic prediction is 
also comparable to that seen in the N-body simulations.  We use 
this fact to argue that our scheme provides MMP histories that are 
similar to those in the numerical simulations.  

Fig.~\ref{zform} shows the formation time distribution obtained 
directly from the numerical Monte-Carlo code.  
Since the distribution is in reasonable agreement with that found 
in N-body simulations, it may be worth studying our algorithm 
further.  Below, we show how to derive explicit analytic expressions 
for the distribution of formation times and masses.  

\subsection{Analytic calculation of the formation time distribution}
Suppose a random variable $Y$ is drawn from a Gaussian distribution
with mean zero and variance $\chi$.
Then $Y^2$ has a Gamma(${1\over 2},\chi$) distribution, 
and $X=Y^{-2}$ is said to have the Inverse Gamma(${1\over 2},\chi$) 
distribution.  That is, 
\begin{equation}
p(X=x)\,{\rm d}x = {(\chi/x)^{1/2}\over \sqrt{2\pi}}\ 
\exp\left(-{\chi\over 2x}\right)\ {{\rm d}x\over x}.
\end{equation}
Consider a sequence of random variables $X_1, X_2,\cdots$, 
with each of the $X_i$s chosen from an 
Inverse Gamma(${1\over2},\chi$) distribution independently of 
the others.  Define 
\begin{equation}
S_n = \sum_{i=1}^n X_n.
\end{equation}
It is easy to verify that the distribution of $S_n$ is 
Inverse Gamma(${1\over 2},n^2\chi$), so that 
\begin{equation}
p_n(S_n=x)\,{\rm d}x = {(n^2\chi/x)^{1/2}\over \sqrt{2\pi}}\ 
\exp\left(-{n^2\chi\over 2x}\right)\ {{\rm d}x\over x}.
\label{pnsnig}
\end{equation}
These expressions will be useful below.  

Consider a halo that has mass $M_0$ at $z_0$.  Suppose that 
at $z_1 = z_0 + \Delta z$, the MMP of this halo has mass 
$0\le M_1\le M_0$ with probability $f(M_1,z_1|M_0,z_0)$, 
where $f(1|0)$ is given by equation~(\ref{ave-num}).  
Note that, in the main text, we use a different function 
for the MMP probability.  We use this form here to 
illustrate the way the calculation is done, and show the result 
of using our rule later.  Then 
\begin{equation}
(S_1 - S_0) = (\delta_{{\rm c}0}\,\Delta z)^2/y^2, 
\end{equation}
where $y$ is drawn from a Gaussian distribution with zero mean 
and unit variance, so $y^2$ is drawn 
from a Gamma(${1\over 2}$,1) distribution, which means that $1/y^2$ 
is drawn from an Inverse Gamma(${1\over 2},1$) distribution.  Thus, 
the distribution of $(S_1-S_0)$ is Inverse Gamma(${1\over 2},\chi$), 
with $\chi=(\delta_{{\rm c}0}\,\Delta z)^2$.  

Suppose that, at $z_2=z_1+\Delta z$, the MMP of $M_1$ 
is obtained by drawing $M_2$ from $f(2|1)$, 
at $z_3=z_2+\Delta z$ the MMP of $M_2$ is drawn from $f(3|2)$, 
and so on.  Then, after $n$ steps, 
\begin{equation}
(S_n - S_0) = (\delta_{{\rm c}0}\,\Delta z)^2\ \sum_{i=1}^n \, y^{-2}, 
\end{equation}
and the distribution of $S_n$ is the distribution of the sum of 
$n$ Inverse Gamma(${1\over 2},\chi$) variates.  Therefore, 
$p_n(S_n-S_0 = x)\,{\rm d}x$ is given by equation~(\ref{pnsnig}), 
with the obvious notation that 
\begin{equation}
n^2\chi = n^2\,(\delta_{{\rm c}0}\,\Delta z)^2 
= \delta_{{\rm c}0}^2\,(z_n-z_0)^2 .
\end{equation}

Define the formation time $t_{\rm f}$ as the first time that 
the mass of the MMP becomes less than $fM_0$, with 
$0.5\le f\le 1$.  Let $S_{\rm f} = \sigma^2(fM_0)$, and let 
$z_{\rm f}$ denote the redshift associated with the formation time.  
Suppose that $\Delta z$ is chosen sufficiently small that 
$z_{\rm f}-z_0 = n\,\Delta z$, with $n$ an integer.  Then, for 
this MMP generating rule, the formation time distribution is got 
from 
\begin{eqnarray}
P(t_{\rm f}<t|M_0,t_0) &=& P(S_n-S_0 < S_{\rm f}-S_0) \nonumber \\
&=& \int_0^{S_{\rm f}-S_0} p_n(x)\,{\rm d}x \nonumber \\
&=& {\rm erfc}
\left({\delta_{{\rm c}0}(z_{\rm f}-z_0)\over\sqrt{2(S_{\rm f}-S_0)}}\right).
\end{eqnarray}
This is the same as equation~(2.23) in Lacey \& Cole (1993).  

Suppose instead that the MMP is not chosen according to $f(1|0)$, 
but it is chosen using $N(1|0) = (M_0/M_1)\,f(1|0)$, with 
the requirement that $M_0/2\le M_1\le M_0$.  This is the choice 
we make in the main text.  (Recall that, as 
$\Delta z = (z_1-z_0)\to 0$, the integral of $N(1|0)\to 1$.)
Suppose further that at $z_2=z_1+\Delta z$ the MMP of $M_1$ is 
chosen from $N(2|1)$, with $M_1/2\le M_2\le M_1$.  
Then, 
\begin{equation}
p(S_2) = \int_{S_{\rm min}}^{S_2} N(2|1)\,N(1|0)\,{\rm d}S_1,
\end{equation}
where $S_{\rm min} = S_0$ if $S_2<S(M_0/2)$, and 
$S_{\rm min} = S(2M_2)$ otherwise.  These restrictions on 
$S_{\rm min}$ follow from the requirements that, in any step, 
the mass of the MMP may not decrease to less than half the current 
mass, and the mass of the MMP can never be greater than $M_0$.  
Now, if $S_2<S(M_0/2)$, then this is the same convolution integral 
as the split rule by $f(1|0)$ above, since the extra factors 
$(M_1/M_2)(M_0/M_1) = (M_0/M_2)$ may be taken outside the integral.  
Thus, provided that $M_2\ge M_0/2$, 
\begin{equation}
p(S_2) = \left({M_0\over M_2}\right)\,f(2|0) = N(2|0).
\end{equation}
Choose some $0.5\le f\le 1$.  If $M_i\ge M_0/2$ for $1\le i\le n$, 
then after $n$ steps, $p(S_n) = N(S_n|S_0)$, and 
\begin{equation}
P(t_{\rm f}<t|M_0,t_0) =
\int_{fM_0}^{M_0} N(m,z_{\rm f}|M_0,z_0)\,{\rm d}m ,
\end{equation}
where $z_{\rm f}-z_0=n\,\Delta z$ as before.  This expression is 
the same as equation~(2.26) in Lacey \& Cole (1993).  
This demonstrates that, for sufficiently small steps in $\Delta z$, 
choosing the MMP using $N(1|0)$ of equation~(\ref{ave-num}) gives 
the correct distribution of formation times (equation~\ref{form}).  

\subsection{Analytic calculation of the distribution of formation masses}
Suppose that the formation time is defined as above for 
some $0.5\le f\le 1$.  Then the formation mass $M_{\rm f}$ may 
be defined in at least two ways.  
The first is to define it as the smallest mass $M$ for 
which $M > fM_0$, averaged over all merger histories.  
This is the same definition as that shown in Figure~11 of 
Lacey \& Cole (1993).  
Alternatively, we could define the formation mass as the 
largest mass $M$ at which $M<fM_0$, averaged over all 
histories.  Compared to the quantity in the first definition, 
this is the mass of the MMP in the next time step.  
For either definition, in the appropriate units, the distribution 
of formation masses for a white-noise spectrum (initial slope $n=0$) 
provides an excellent fit to the distribution for other power spectra.
(Recall that the same is true for the formation time distribution.)  
This is fortunate, since, when $n=0$, then, for our MMP generating 
algorithm, the distribution of formation masses can be calculated 
analytically.  We show this below.  

As above, it is convenient to work in terms of the variable 
$S$ rather than $M$.  The quantity described by the first 
definition can be computed as follows. 
The probability that $M_{\rm f}=M$ is equal to the probability 
that after step $n$ the mass was $M_n=M>fM_0$, and that the mass 
dropped to $M_{n+1}<fM_0$ on step $n+1$, integrated over all 
allowed values of $M_{n+1}$, and summed over all values of $n$.  
The sum over all $n$ allows for all possible formation times. 
Thus, 
\begin{eqnarray}
p(M)\,{\rm d}M &=& p(S)\,{\rm d}S \nonumber \\
&=& {\rm d}S\ \sum_{{\rm allowed}\ n} \int_{S_{\rm f}}^{S(M/2)}
p(S_{n+1}|S_n=S)\,p(S_n=S|S_0)\,{\rm d}S_{n+1} ,
\end{eqnarray}
where $S_{\rm f} = S(fM_0)$.  The upper and lower limits to the 
integral reflect the fact that the allowed values of $M_{n+1}$ 
are all those that are smaller than $fM_0$ but larger than 
$M_n/2$.  Now, the order of the sum and the integral can be 
rearranged.  Since $p(S_{n+1}|S_n=S)$ is just the one-step 
split rule, it is independent of $n$, so it is convenient to do 
the sum over $n$ first.  Since $S_n$ is less than 
$S_{\rm f}=S(fM_0)$, the previous section showed that, for our 
algorithm, $p(S_n|S_0)\,{\rm d}S_n = N(S_n,z_n|S_0,z_0)$. 
Moreover, for small $\Delta z$, summing over $n$ is like 
integrating over all formation times $z_{\rm f}$.  Thus, 
\begin{eqnarray}
{\rm d}S \,\sum_n p(S|S_0) &\to& 
{{\rm d}S\over \Delta z}\,\int_0^\infty {\rm d}z_{\rm f}\,
N(S,z_{\rm f}|S_0,z_0) \nonumber \\
&=& {{\rm d}S\over (S-S_0)}\,{(M_0/M)\over \Delta z}
\sqrt{S-S_0\over 2\pi\delta^2_{\rm c0}} .
\end{eqnarray}
This expression is independent of $M_{n+1}$.  
For a white-noise power-spectrum, $M/M_{n+1} = S_{n+1}/S$, so 
the integral over the allowed range of $S_{n+1}$ can be done 
analytically:  
\begin{equation}
\int_{S_{\rm f}}^{S(M/2)}
N(S_{n+1}|S_n=S)\,{\rm d}S_{n+1} = 
2 \sqrt{D\over\pi}
\left({\rm e}^{-D} - {{\rm e}^{-Dq}\over \sqrt{q}}\right) + 
(1-2D)\,\Bigl({\rm erf}(\sqrt{Dq}) - {\rm erf}(\sqrt{D})\Bigr)
\end{equation}
where $D = (\delta_{\rm c0}\,\Delta z)^2/2S$, and 
$q = S/(S_{\rm f}-S)$.  
Thus, as $\Delta z\to 0$, so $D\to 0$, then, provided $Dq\ll 1$, 
\begin{equation}
p(s)\,{\rm d}s \to {1\over m\pi}
{{\rm d}s\over \sqrt{(s-1)(s_{\rm f}-s)}}
\left(2 - {s_{\rm f}\over s}\right),\qquad {\rm for}\ n=0, 
\label{mfmore}
\end{equation}
where $s = S/S_0$, $s_{\rm f} = S_{\rm f}/S_0$, and $m=M/M_0$.
In this limit, the dependence on $\Delta z$ has dropped out.  

\begin{figure}
\centering
\mbox{\psfig{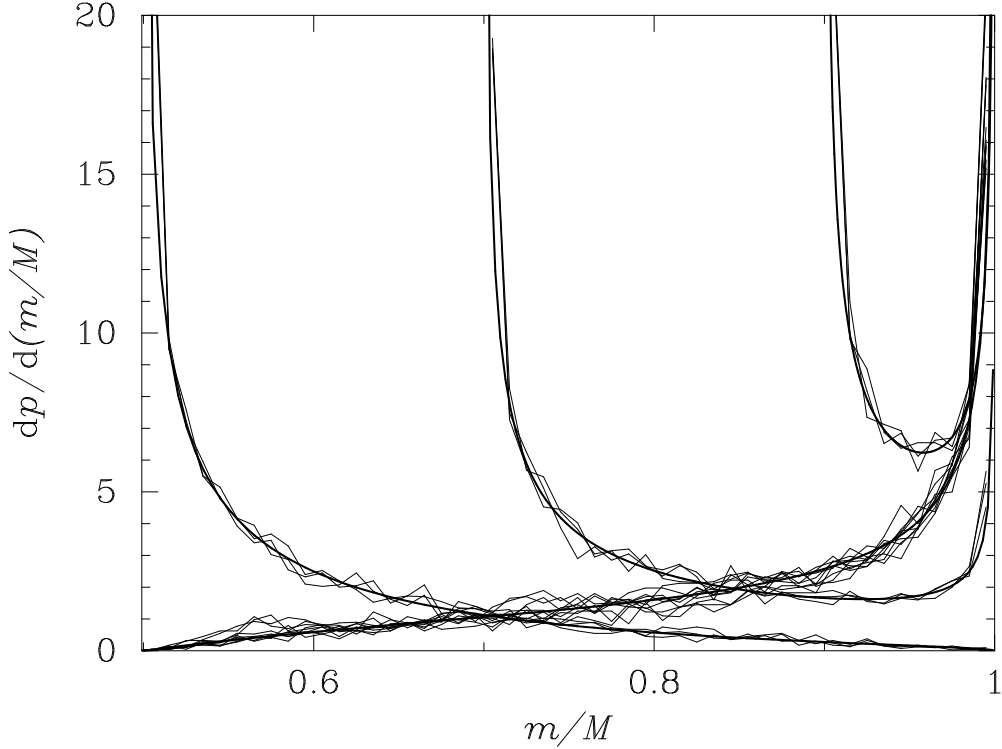}}
\caption{Distribution of formation masses using our MMP generating
algorithm.  The thick solid curve that is lowest on the left shows 
equation~(\ref{mflim}) with $f=0.5$ and $M=fM_0$.  Curves for 
other values of $f$ are similar so we have not shown them.  
The other thick curves show equation~(\ref{mfmore}) with $f=0.5$, 
$0.7$ and $0.9$, and $M=M_0$.  
The thin curves show the corresponding quantities computed using 
our Monte-Carlo algorithm for a range of values for the spectral 
slope $n$ (0, $-1$ and $-2$), the final mass $M_0$ ($0.1$, $1$, 
and $10M_*$), and the time step $\Delta z$ (0.01--0.1).  
All the curves are very similar, and they are well fit by the 
analytic (white-noise) formulae.}
\label{emform}
\end{figure}

The quantity described by the second definition can be computed 
similarly.  The probability that $M_{\rm f}=M$ is 
equal to the probability that after step $n$ the mass was 
$M_n>fM_0$, and that the mass dropped to $M<fM_0$ on step $n+1$,   
integrated over all allowed values of $M_n$, and summed over 
all choices of $n$.  That is, 
\begin{eqnarray}
p(M)\,{\rm d}M &=& p(S)\,{\rm d}S \nonumber \\
&=& {\rm d}S\ \sum_{{\rm allowed}\ n} \int_{S{\rm min}}^{S_{\rm f}}
p(S_{n+1}=S|S_n)\,p(S_n)\,{\rm d}S_n ,
\end{eqnarray}
where $S_{\rm min} = S(2M)$ if $2M<M_0$, 
and $S_{\rm min} = S_0$ otherwise because, for our algorithm, 
$M_{\rm f}=M$ must lie in the range $fM_0/2\le M\le fM_0$. 
This means that $M_n$ lies in the range $M\le M_n\le 2M$, 
unless $2M>M_0$, in which case the maximum allowed 
value of $M_n$ is simply $M_0$.  The sum is over all formation 
times, so it is over all $n>0$.  Now, as before, 
$p(S_n)\,{\rm d}S_n = N(S_n,z_n|S_0,z_0)$, and 
$p(S_{n+1}=S|S_n)$ is just the one step split rule, so 
it is independent of $n$.  If we rearrange the order of the sum 
and the integral above, then, for small $\Delta z$, summing over 
$n$ is like integrating over all formation times $z_{\rm f}$, 
so the integral is the same as before:    
\begin{eqnarray}
{\rm d}S_n \,\sum_n p(S_n) &\to& 
{{\rm d}S_n\over \Delta z}\,\int_0^\infty {\rm d}z_{\rm f}\,
N(S_n,z_{\rm f}|S_0,z_0) \nonumber \\
&=& {{\rm d}S_n\over (S_n-S_0)}\,{(M_0/M_n)\over \Delta z}
\sqrt{S_n-S_0\over 2\pi\delta^2_{\rm c0}} .
\end{eqnarray}
Using this expression and then doing the integral over the 
allowed range of $S_n$ implies that 
\begin{equation}
p(S)\,{\rm d}S = {{\rm d}S\over S-S_0}
\left({M_0\over M}\right)\, {{\rm e}^{-A}\over \sqrt{A}}\, 
\left[{\rm erf}(\sqrt{AB}) - {\rm erf}(\sqrt{AC})\right]
\label{mfless}
\end{equation}
where 
\begin{displaymath}
A = {(\delta_{\rm c0}\,\Delta z)^2\over 2 (S-S_0)}, \quad
B = \left({S_{\rm f}-S_0\over S-S_{\rm f}}\right),\quad{\rm and}\ 
C = {\rm max}\left(0,1-{S_0\over S_{\rm min}}\right) .
\end{displaymath}
For small $\Delta z$, the error function terms become 
$2 \sqrt{(A/\pi)} (\sqrt{B} - \sqrt{C})$
provided $(S-S_{\rm f})/S_{\rm f}$ is not small.  
In this approximation, the formation mass distribution is 
\begin{eqnarray}
p(s)\,{\rm d}s &=& 
\left({M_0\over M}\right) {1\over \pi (s-s_0)} 
\,\sqrt{1-s_0\over s-1}\,{\rm d}s \qquad\qquad\qquad\qquad
{\rm if}\ 2M>M_0, \nonumber \\
&=& \left({M_0\over M}\right) {1\over \pi (s-s_0)} 
\left[\sqrt{1-s_0\over s-1} - \sqrt{1 - {s_0\over s_{\rm min}}}\right]
\,{\rm d}s \quad {\rm if}\ M_0\ge 2M\ge fM_0, 
\label{mflim}
\end{eqnarray}
where $s_i \equiv S_i/S_{\rm f}$.  Notice that, in this limit, 
the dependence on $\Delta z$ has dropped out.  

As with the distribution of formation times, this is a 
well-behaved probability distribution only for white-noise, where 
$s_i = S_i/S_{\rm f} = fM_0/M_i$.  However, as with 
the formation time distribution, this distribution depends only 
very weakly on power spectrum, so the white-noise expression 
provides a good fit to the distribution for other power-spectra.  
It also turns out that this distribution depends only weakly on 
$f$, so, in Figure~\ref{emform}, only the curve for $f=0.5$ is 
shown.  

Figure~\ref{emform} shows both formation mass distributions for a 
range of initial power-spectra, final $M_0$s, and choices for $f$.  
The $x$-axis shows $M_{\rm f}/M_0$ for equation~(\ref{mfmore}), 
and $M_{\rm f}/(fM_0)$ for equation~(\ref{mfless}).  
The thick solid curves show the analytic formulae 
with $f=0.5$, $0.7$ and $0.9$, for $M/M_* = 0.1$, $1$, and $10$, 
and $n=0$.  The thinner lines show the associated distributions 
obtained using our MMP generating code for 
$n=0$ (solid), $n=-1$ (dashed), and $n=-2$ (dotted).  
The curves for the different power-spectra are virtually 
indistinguishable, and they are all well fit by our analytical 
formulae.  This, and the fact that our code also generates the 
correct distribution of formation times (Fig.~1), shows that 
our MMP generating code works.  


\bigskip
\begin{thebibliography}{}
\def\ref{\par\noindent\hangindent 15pt} 
\def\apj#1{{\it Astrophys. J.}
{ #1}} 
\def\apjl#1{{\it Astrophys. J. (Lett.)} { #1}} 
\def\apjs#1{{\it
Astrophys. J. (suppl.)} { #1}} \def\mn#1{{\it M.N.R.A.S.} { #1}}
\def\aa#1{{\it Astron. Astrophys.} { #1}} 
\def\nat#1{{\it Nature} {
#1}} 
\def\ana#1{{\it Annu. Rev. Astron. Astrophys.} { #1}}

\bibitem{} Avila-Reese V., Firmani C., Hern\'andez X. 1997, astro-ph/9710201
\bibitem{} Bertschinger, E. 1985, \apjs{58}, 39
\bibitem{} Bond, J.R., Kaiser, N., Cole, S. $\&$ Efstathiou, G. 1991,
\apj{379}, 440
\bibitem{} Bower, R.G. 1991, \mn{248},332
\bibitem{} Chandrasekhar, S. 1943, \apj{97}, 255
\bibitem{} Dekel, A. 1981, \aa{101}, 79
\bibitem{} Dubinski, J \& Calberg, R. 1991, \apj{378}, 496
\bibitem{} Eisenstein, D.J. \& Loeb, A. 1996, \apj{459}, 432
\bibitem{} Filmore, J.A. $\&$ Goldreich, P. 1984, \apj{281}, 1
\bibitem{} Gunn, J. $\&$ Gott, J.R. 1972, \apj{179},1
\bibitem{} Hoffman, Y. $\&$ Shaham, J. 1985, \apj{297}, 16
\bibitem{} Hoffman, Y. 1988, \apj{i328}, 489
\bibitem{} Huss A., Jain B., Steinmetz M.  1998, astro-ph/9803117
\bibitem{} Landau, L.D. $\&$ Lifshitz, E.M. 1960, {\it Mechanics},
(course of theoretical physics; V.1), Pergamon Press
\bibitem{} Kravtsov, A.V., Klypin, A.A., Bullock, J.S. \& Primack,
J.R.
1997, astro-ph/9708176
\bibitem{} Lemson, G. 1995, PhD. thesis, RIJKSUNIVERSITEIT GRONINGEN
\bibitem{} Lynden-Bell, D. 1967, \mn{136}, 101 
\bibitem{} Moore, B., Chigna, S., Governato, F., Lake, G., Quinn, T. 
$\&$ Stadel, J. 1997, astro-ph/9711259
\bibitem{} Navarro, J.F., Frenk, C.S. $\&$ White, S.D.M. 1996, \apj{462}, 563
\bibitem{} Nusser, A.  $\&$ Dekel, A. 1992, \apj{362}, 14
\bibitem{} Ostriker, J.P. \& Turner, E.L. 1979, \apj{234}, 785
\bibitem{} Peebles, P. J. E. 1989, \apjl{344}, 53
\bibitem{} Sheth, R.K. 1996, \mn{291}, 1277
\bibitem{} Sheth, R.K. $\&$ Pitman, J. 1997, \mn{289}, 66
\bibitem{} Syer, D. $\&$ White, S.D.M. 1996, astro-ph/9611065
\bibitem{} White, S.D.M. 1976, \mn{177}, 717
\bibitem{} White S.D.M., Zaritsky D.  1992, \apj{394},1
\bibitem{} White, S.D.M. 1983, \apj{274},53
\bibitem{} White, S.D.M. 1996, astro-ph/9602021
\bibitem{} Zhao, H.S. 1996, \mn{278}, 488

\end{thebibliography}
\end{document}